\documentstyle[11pt,moriond,epsfig]{article}
\def\be{\begin{equation}}
\def\ee{\end{equation}}
\def\bea{\begin{eqnarray}}
\def\eea{\end{eqnarray}}

\begin{document}
\title{FINITE TEMPERATURE TRANSPORT IN INTEGRABLE\\
QUANTUM MANY BODY SYSTEMS
\footnote{to be published in Proceedings of XXXIV Rencontres
de Moriond Conference on ``Quantum Physics at Mesoscopic Scale", France, 
January 1999}
}

\author{X. ZOTOS, F. NAEF }

\address{Institut Romand de Recherche Num\'erique en Physique 
des Mat\'eriaux,\\(EPFL-PPH) Ecublens, 1015 Lausanne, Switzerland}

\author{P. PRELOVSEK}

\address{ Faculty of Mathematics and Physics,\\ and J. Stefan Institute,
1000 Ljubljana, Slovenia}

\maketitle\abstracts{
Recent developments in the analysis of finite temperature 
dissipationless transport in integrable quantum many body problems are 
presented. In particular, we will discuss: (i) the role played by the 
conservation laws in systems as the spin 1/2 Heisenberg chain and the 
one-dimensional Hubbard model, (ii) exact results obtained using the Bethe
ansatz method on the long time decay of current correlations.}

\section{Introduction}
It has recently being proposed that integrable quantum many body systems 
show dissipationless transport at finite temperatures \cite{m96}. 
This idea was motivated by analytical and numerical studies on 
a toy model \cite{czp} and the fermionic version of the 
Heisenberg model \cite{zp}.
It is analogous to the well known effect of transport by 
solitons in classical nonlinear integrable systems. In this domain of 
classical physics, a part of the activity is already in the  
technological applications, for instance the propagation of solitons 
in optical fibers \cite{optfib} and in particular the robustness of 
dissipationless transport to perturbations.

In quantum many body systems, there is not yet clear experimental evidence 
for unusually high conductivity (close to ideal) at finite temperatures 
for a system described by a (nearly) integrable Hamiltonian. Perhaps 
the best candidates so far are quasi one dimensional spin 1/2 systems 
described by the Heisenberg model. In fact NMR studies \cite{tak} on the 
$Sr_2CuO_3$ compound showed an unusually high value of the diffusion constant, 
characteristic of ballistic rather than diffusive behavior. 

Besides the magnetic compounds, systems characterized by strong short range 
electronic correlations (described by the integrable one dimensional 
Hubbard model) could be candidates for observing the predicted 
ballistic behavior, even more conspicuous at high temperatures, instead 
of the expected diffusive one due to electron-electron scattering. 
Nanotubes or artificially made nanostructures come to mind as possible 
experimental realizations.

From the experimental point of view, the most relevant question is the 
robustness of this dissipationless transport to perturbations,  
for instance impurities, 3d coupling, phonons. 
The same problem analyzed for classical systems, indicates relative 
insensitivity of the soliton propagation to disorder \cite{optfib}. 
From the theoretical perspective, 
an analytical solution of the dynamic properties of integrable systems 
should be expected, exactly due to the integrability of the models.
On the other hand, the analysis of robustness might 
largely depend on numerical simulation studies, the same situation as in 
classical systems. We should also point out, that this problem is closely 
related to studies on dynamical systems, analyzing the conditions of 
appearance and stability of classical/quantum chaos \cite{jl,pros}.  

The framework for discussing dynamic properties at finite 
temperatures/frequencies is the linear response theory (or Kubo formalism). 
In this formulation, using the fluctuation - dissipation theorem, the 
transport properties are related to the study of dynamic correlations at 
equilibrium. It might be objected that results drawn from this method for 
integrable systems are questionable. However, we can point out that 
transport properties derived by linear response theory for noninteracting 
systems (e.g. an harmonic crystal) \cite{leb} or classical integrable 
systems \cite{livi} are in accord with experimental observations \cite{beck} 
and results from more sophisticated theoretical methods.

The basic way for characterizing the transport properties is by 
the dynamic current correlations $<j(t)j>$. As a criterion for 
dissipationless transport we will study the long 
time asymptotic value $C_{jj}$, $<j(t)j(0)>_{t\rightarrow \infty}=C_{jj}$.
If $C_{jj}$ is finite, the response is reactive and the system shows ideal 
conducting behavior.
$C_{jj}$ is related to the Drude weight $D(T)$ in the real part of the 
conductivity as expressed in linear response theory:
\be
\sigma(\omega)=2\pi D\delta(\omega)+\sigma_{reg}(\omega),~~~~
D=\frac{\beta}{2L} C_{jj}
\ee

It should be noted that a vanishing $D(T)$ does not necessarily imply 
diffusive behavior. If the current-current correlations decay too slowly 
then transport coefficients cannot be defined. 
In the analogous situation of one dimensional classical systems, different  
behaviors have been observed \cite{livi}; typically if a system is integrable 
then $C_{jj}>0$ and the system is an ideal conductor. If it is nonintegrable, 
then models have been found where correlations decay fast enough 
and a transport coefficient can be defined, but there 
are also models where the decay is too slow indicating anomalous low 
frequency conductivity. In quantum many body systems the situation is less 
clear, but we might similarly expect a variety of behaviors.

In the following we will discuss two recent developments for analyzing the 
transport properties in integrable systems: the first, it is the way 
in which the conservation laws characterizing these systems affect their 
long time current correlations \cite{znp}; the second, it is a 
procedure \cite{fk} based on the Bethe ansatz method for calculating 
analytically $D(T)$. 

\section{The role of conservation laws}

Integrable many body systems are characterized by
a macroscopic number of conserved quantities \cite{lu}.
A set of conservation laws is represented by local involutive operators $Q_n$,
commuting with each other $[Q_n,Q_m]=0$ and with the Hamiltonian, $[Q_n,H]=0$.
We can relate the time decay of correlations to local conserved
quantities in Hamiltonian systems by using an inequality proposed
by Mazur \cite{maz}

\be
\lim_{T\rightarrow \infty} \frac{1}{T} \int_0^T <A(t)A> dt \geq \sum_n
\frac{<A Q_n>^2}{<Q_n^2>}
\label{mazur}
\ee
Here $< >$ denotes thermodynamic average,
the sum is over a subset of conserved quantities ${Q_n}$, orthogonal
to each other $<Q_n Q_m>=<Q_n^2>\delta_{n,m}$,
$A^{\dagger}=A$ and $<A>=0$.
Applying this inequality to current-current correlations we obtain: 
\be
D \geq \frac{\beta}{2L} \sum_n \frac{<j Q_n>^2}{<Q_n^2>} .
\label{eqd}
\ee

We will now discuss two applications of this inequality,   
the decay of spin currents in the Heisenberg model and the 
electrical conductivity in the Hubbard model \cite{znp}.

\subsection{s=1/2 Heisenberg model}
It is described by the Hamiltonian:
\be
H=\sum_{l=1}^L (J_x S_l^x S_{l+1}^x +
J_y S_l^y S_{l+1}^y + J_z S_l^z S_{l+1}^z)
\label{heis}
\ee

\noindent
(by a Jordan-Wigner transformation the Heisenberg model is
equivalent to a model of spinless fermions interacting with a
nearest-neighbor interaction ($t-V$ model)).

This model is characterized by a macroscopic number of conservation laws.
It is worth to note that the first nontrivial conservation law $Q_3$ is 
related to a physical quantity, it coincides with the energy current $j^E$:

\be
Q_3=j^E=\sum_{l=1}^L J_x J_y
(S_{l-1}^x S_l^z S_{l+1}^y -
S_{l-1}^y S_l^z S_{l+1}^x) + (x,y,z).
\label{hje}
\ee
This is particularly interesting and of actual experimental 
relevance \cite{kom} 
as it implies ideal thermal conductivity (energy currents do not decay at 
all, $<j^E(t)j^E>=const.$). It should also be taken into account in the 
analysis of the quasielastic peak in Raman experiments where the local dynamic 
energy correlations enter \cite{kur,nz}.

Regarding the spin current correlations, for $J_x=J_y=1, J_z=\Delta$ we find 
that: 

\be
D=\frac{\beta}{2L} C_{j^zj^z}
\geq \frac{\beta}{2L}\frac{<j^z Q_3>^2}{<Q_3^2>}
\label{dz}
\ee
where $j^z=$spin current$=\sum_{l=1}^L (S_l^y S_{l+1}^x-S_l^x S_{l+1}^y)$,
implying ideal spin conductivity and the absence of spin diffusion. 
We can analytically calculate the right hand size of this inequality in the 
$\beta \rightarrow 0$ limit,

\be
D\geq \frac{\beta}{2} \frac{8 \Delta^2 m^2 (1/4-m^2)}
{1+8\Delta^2(1/4+m^2)},~~~~~~~m=<S_l^z>
\label{dz0}
\ee
From this result we notice that we obtain a positive bound only for 
a finite magnetization $m$, when the system is in a magnetic field.
For $m=0$ this inequality does not provide a useful bound although, as we 
will show below using the Bethe ansatz method, $D(T)$ is still positive.

\subsection{Hubbard model}

Finally, for the Hubbard model described by the Hamiltonian,

\be
H=(-t) \sum_{\sigma,i=1}^L (c_{i\sigma}^{\dagger} c_{i+1 \sigma} + h.c.)
+ U \sum_{i=1}^L (n_{i\uparrow}-\frac{1}{2})(n_{i\downarrow}-\frac{1}{2})
\label{hhubb}
\ee
considering $Q_3$ given by:
\be
Q_3=\sum_{\sigma} (-t)^2
(i c_{i+1\sigma}^{\dagger} c_{i-1\sigma} + h.c.)-
U (j_{i-1,i,\sigma}+j_{i,i+1,\sigma})
(n_{i,-\sigma}-\frac{1}{2})
\label{hq3}
\ee
we obtain for the decay of current correlations in the 
$\beta \rightarrow 0$ limit,
\be
D \geq \frac{\beta}{2} \frac
{[U\sum_{\sigma}2\rho_{\sigma}(1-\rho_{\sigma})(2\rho_{-\sigma}-1)]^2}
{\sum_{\sigma} 2\rho_{\sigma}(1-\rho_{\sigma})[1+U^2(2\rho_{-\sigma}^2-
2\rho_{-\sigma}+1)]}
\label{dhubb}
\ee

Again, we obtain a finite value for $D(T)$ and so ideal conducting behavior 
for a system out of half-filling. Bethe ansatz analysis by Fujimoto and 
Kawakami \cite{fk} confirm this result (the system at half-filling has also 
been analyzed).

\section{Bethe ansatz analysis for the Heisenberg model}

Recently a new procedure was proposed for analytically calculating $D(T)$ 
using the Bethe ansatz method \cite{fk}.The application to the Heisenberg 
spin 1/2 model, that we will now sketch \cite{xz}, is based on a series of 
assumptions and developments.

First, the structure of the solutions of the Bethe ansatz equations 
for $0\leq\Delta\leq 1$ conjectured by Takahashi and Suzuki \cite{ts} 
is used.  Parametrizing the anisotropy coupling by 
$\Delta=\cos\theta,\;\;\theta=\frac{\pi}{\nu},\;\; \nu \mbox{  integer}$,
and the pseudomomenta $k_{\alpha}$ by the rapidities $x_{\alpha}$, 
$\cot(\frac{k_{\alpha}}{2})=\cot(\frac{\theta}{2})
\tanh(\frac{\theta x_{\alpha}}{2})$, the solutions are organized into 
a finite number of allowed strings \cite{str}, $n=1,2,...,\nu-1$,  

\begin{eqnarray}
x_{\alpha,+}^{n,k}&=&x_{\alpha}^n+(n+1-2k)i+O(e^{-\delta N});~~~k=1,2,...n\\
x_{\alpha,-}&=&x_{\alpha}+i\nu+O(e^{-\delta N}),~~~\delta > 0.
\end{eqnarray}

Next, finite size corrections on the positions of the strings 
are expressed \cite{bm} by introducing the functions $g_{1j}$ and $g_{2j}$:

\be
x_N^j=x_{\infty}^j+\frac{g_{1j}}{N}+\frac{g_{2j}}{N^2}
\ee
where $x_N^j(x_{\infty}^j)$ are the rapidities for a system of
size $N(\infty)$.

Finally introducing excitation and hole densities $\rho_j$ and $\rho_j^h$
as in the standard thermodynamic Bethe Ansatz and expanding the Bethe ansatz 
equations to $O(\frac{1}{N})$ and $O(\frac{1}{N^2})$ we obtain:

\be
D=\frac{1}{2}\beta\sum_j\int_{-\infty}^{+\infty}dx (\rho_j+\rho_j^h)
<n_j>(1-<n_j>) (\frac{\partial \epsilon_j}{\partial x}
\frac{\partial g_{1j}}{\partial\phi})_{|\phi\rightarrow 0}
\ee
where $<n_j>=1/(1+e^{\beta\epsilon_j})$ and
$\epsilon_j=(1/\beta)\ln(\rho_j^h/\rho_j)$.
This expression looks formally analogous to that for free fermions:
\be
D=\frac{\beta}{2N}\sum_{\mu} <n_{\mu}>(1-<n_{\mu}>)
(\frac{\partial \epsilon_{\mu}}{\partial\phi})^2_{|\phi\rightarrow 0}
\ee
where $<n_{\mu}>$ is the Fermi-Dirac distribution and $\epsilon_{\mu}$ are 
single particle energies.

The functions $\rho_j$, $\rho_j^h$, $g_{1j}$, $g_{2j}$
are determined numerically by iteration.
Using this procedure we obtain $D(T)$ for all temperatures and 
couplings $0\leq\Delta\leq 1$ as is shown in Fig. 1. 

\begin{figure}[ht]
\epsfxsize 8cm \centerline{\epsffile{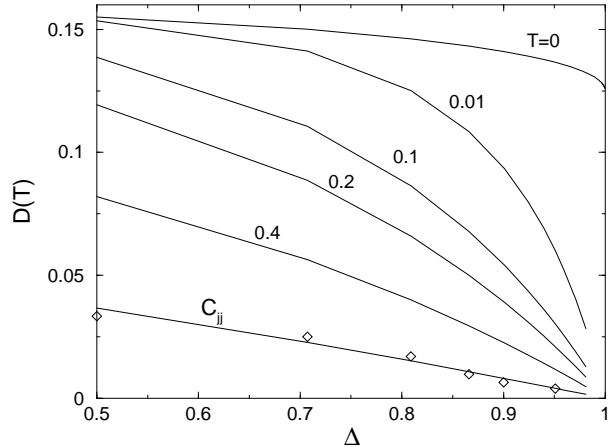}}
\caption{
$D(\Delta)$ evaluated at the points $\nu=3,...,16$ and various
temperatures. The continuous line
is the high temperature proportionality constant $C_{jj}=D/\beta$.
The $\diamond$ indicate exact diagonalization results
from ref. (17) }
\end{figure}

In particular, we recover the known $T=0$ value \cite{ss} 
$D_0$ with a characteristic power law behavior:
\be
D(T)=D_0-const.T^{\alpha},~~~ \alpha=2/(\nu -1), 
\ee
and we obtain good agreement with numerical simulations \cite{nz} 
for $\beta=0$ (diamonds), 
a result that lends support both to the Bethe ansatz procedure and 
the string assumption.

Particularly interesting is the vanishing of $D(T)$ for $\Delta=1$ at any 
finite temperature. This result suggests that for $\Delta\geq 1$ the 
Heisenberg model does not show ideal conducting behavior and the isotropic 
model is a borderline case. If the conjecture,
born out of a numerical study \cite{zp}, of absence of weight in the 
conductivity at low frequencies proves valid in the thermodynamic  limit, 
then we would have a realization of an exotic ideal insulating phase.
Finally, from the results presented, it is not unreasonable to expect  
anomalous low frequency spin dynamics for the isotropic Heisenberg model, 
hopefully observable in experiments.

\section*{Acknowledgments}
This work was supported by the Swiss National Science Foundation grant
No. 20-49486.96, the University of Fribourg and the 
University of Neuch\^atel.

\end{document}